\documentclass[fleqn,usenatbib]{mnras}

\usepackage{newtxtext,newtxmath}

\usepackage[T1]{fontenc}
\usepackage{ae,aecompl}
\usepackage{graphicx}	% Including figure
\usepackage{amsmath}	% Advanced maths 
\usepackage{amssymb}	% Extra maths 
\usepackage{marvosym}
\usepackage{hyperref}
\usepackage{dutchcal}
\usepackage[dvipsnames]{xcolor}
\usepackage{flushend}
\newcommand{\beq}[1]{\begin{equation}\label{#1}}
\newcommand{\eeq}{\end{equation}}
\newcommand{\beqn}[1]{\begin{eqnarray}\label{#1}}
\newcommand{\eeqn}{\end{eqnarray}}
\newcommand{\sub}[1]{_\mathrm{#1}}

\newcommand{\tage}{t\sub{age, \star}}
\newcommand{\Op}{\Omega\sub{p}}
\newcommand{\Os}{\Omega\sub{\star}}
\newcommand{\Obar}{\Bar{\Omega}\sub{\star}}
\newcommand{\Ocri}{\Omega\sub{c}}

\newcommand{\Osini}{\Omega\sub{ini, \star}}

\newcommand{\Rp}{R\sub{p}}
\newcommand{\Rpo}{R\sub{p,o}}

\newcommand{\Mp}{M\sub{p}}
\newcommand{\Mpo}{M\sub{p,o}}
\newcommand{\Mprate}{\dot{M}\sub{p}}

\newcommand{\Mjup}{\mathrm{M}\sub{J}}
\newcommand{\Rjup}{\mathrm{R}\sub{J}}
\newcommand{\Msun}{\mathrm{M}_{\odot}}
\newcommand{\Rsun}{\mathrm{R}_{\odot}}
\newcommand{\Osun}{\Omega_{\odot}}

\newcommand{\Mstar}{M\sub{\star}}
\newcommand{\Mstaro}{M\sub{\star,o}}
\newcommand{\Rstar}{R\sub{\star}}
\newcommand{\Mstarrate}{\dot{M}\sub{\star}}
\newcommand{\Msunrate}{\dot{M}_{\odot}}

\newcommand{\apos}{a\sub{p}}

\newcommand{\npp}{n\sub{p}}

\newcommand{\Kpp}{k\sub{2,p}}
\newcommand{\Qp}{Q\sub{p}}
\newcommand{\Kss}{k\sub{2,\star}}
\newcommand{\Qs}{Q_\star}
\newcommand{\koQp}{\Kpp/\Qp}
\newcommand{\koQs}{\Kss/\Qs}

\newcommand{\Ip}{I\sub{p}}
\newcommand{\Is}{I\sub{\star}}
\newcommand{\Lorb}{L\sub{orb}}
\newcommand{\Der}{\mathrm{d}}

\newcommand{\porb}{P\sub{orb}}
\newcommand{\prot}{P\sub{rot, \star}}

\newcommand{\epp}{\varepsilon\sub{p}}
\newcommand{\eps}{\varepsilon\sub{\star}}
\newcommand{\gyrp}{\zeta\sub{p}}
\newcommand{\gyrs}{\zeta\sub{\star}}
\newcommand{\omp}{\Dot{\Omega}\sub{wind, p}}
\newcommand{\oms}{\Dot{\Omega}\sub{wind, \star}}
\newcommand{\roche}{a\sub{Roche}}

\newcommand{\alphap}{\alpha\sub{p}}
\newcommand{\alphapo}{\alpha\sub{p,o}}
\newcommand{\betap}{\beta\sub{p}}
\newcommand{\betapo}{\beta\sub{p,o}}
\newcommand{\gammap}{\gamma\sub{p}}

\newcommand{\alphas}{\alpha\sub{\star}}
\newcommand{\betas}{\beta\sub{\star}}
\newcommand{\betaso}{\beta\sub{\star,o}}
\newcommand{\gammas}{\gamma\sub{\star}}

\title[The spiral-in of ultra-short period planets]
{The impact of tidal friction evolution on the orbital decay of ultra-short period planets}

\author[Alvarado-Montes et al.]{\parbox{\textwidth}{Jaime A. Alvarado-Montes$^{1, 2}$\thanks{E-mail: jaime-andres.alvarado-montes@hdr.mq.edu.au}, Mario Sucerquia$^{3,4,5}$, Carolina Garc\'ia-Carmona$^5$,\\
Jorge I. Zuluaga$^{5}$, Lee Spitler$^{1, 2}$ and Christian Schwab$^{1, 2}$
}\vspace{0.4cm}\\
$^{1}$Department of Physics \& Astronomy, Macquarie University -- Sydney, NSW 2109, Australia.\\
$^{2}$Centre for Astronomy, Astrophysics and Astrophotonics, Macquarie University -- Sydney, NSW 2109, Australia.\\
$^{3}$ N\'ucleo Milenio de Formaci\'on Planetaria (NPF), Chile. Av. Gran Breta\~na 1111, Valpara\'iso, Chile \\
$^{4}$Instituto de F\'isica y Astronom\'ia, Facultad de Ciencias, Universidad de Valpara\'iso, Av. Gran Breta\~na 1111, 5030 Casilla, Valpara\'iso, Chile\\
$^{5}$SEAP research group, Instituto de F\'{\i}sica, FCEN, Universidad de Antioquia --
Calle 70 No. 52-21, Medell\'in, Colombia.
}%

\date{Accepted XXX. Received YYY; in original form ZZZ}

\pubyear{2021}

\begin{document}
\label{firstpage}
\pagerange{\pageref{firstpage}--\pageref{lastpage}}
\maketitle

\begin{abstract}
Unveiling the fate of ultra-short period (USP) planets may help us understand the qualitative agreement between tidal theory and the observed exoplanet distribution. Nevertheless, due to the time-varying interchange of spin-orbit angular momentum in star-planet systems, the expected amount of tidal friction is unknown and depends on the dissipative properties of stellar and planetary interiors. In this work, we couple structural changes in the star and the planet resulting from the energy released per tidal cycle and simulate the orbital evolution of USP planets and the spin-up produced on their host star. For the first time, we allow the strength of magnetic braking to vary within a model that includes photo-evaporation, drag caused by the stellar wind, stellar mass loss, and stellar wind enhancement due to the in-falling USP planet. We apply our model to the two exoplanets with the shortest periods known to date, NGTS-10b and WASP-19b. We predict they will undergo orbital decay in  time-scales that depend on the evolution of the tidal dissipation reservoir inside the star, as well as the contribution of the stellar convective envelope to the transfer of angular momentum. Contrary to previous work, which predicted mid-transit time shifts of $\sim30-190$ s over 10 years, we found that such changes would be smaller than 10 s. We note this is sensitive to the assumptions about the dissipative properties of the system. Our results have important implications for the search for observational evidence of orbital decay in USP planets, using present and future observational campaigns.

\end{abstract}

% Select between one and six entries from the list of approved keywords.
% Don't make up new ones.

%%%%%%%%%%%%%%%%%%%%%%%%%%%%%%%%%%%%%%%%%%%%%%%%%%

%%%%%%%%%%%%%%%%% BODY OF PAPER %%%%%%%%%%%%%%%%%%

\section{Introduction}
\label{sec:intro}

\begin{figure}
        \hspace{-11pt}
        \includegraphics[scale=0.685]{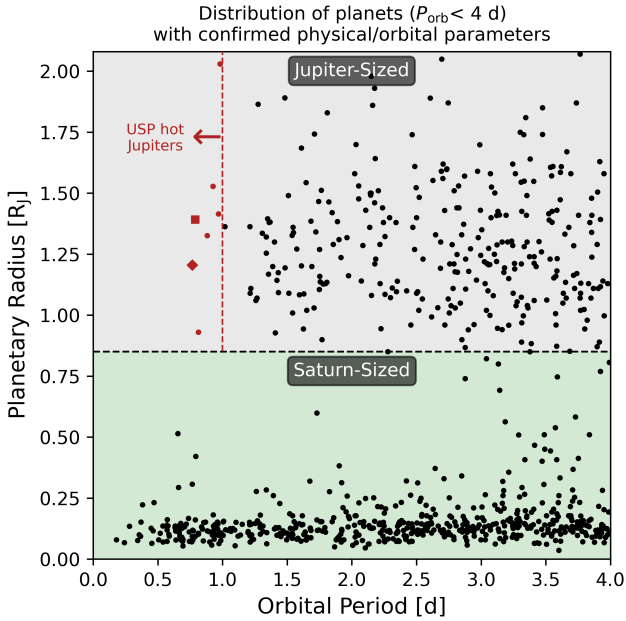}
    \caption{Planetary radius and orbital period of confirmed Jupiter-sized (grey region) and Saturn-sized (green region) close-in exoplanets with $P\sub{orb}\lesssim 4$ d. Planets in the grey region whose $P\sub{orb}\lesssim1$ d (dashed red line) are dubbed ultra-short period (USP) hot Jupiters. Most close-in planets in the green region are below $0.5\rm\,R_{J}$ and hence a desert seems to lie between these two populations. The position of NGTS-10b and WASP-19b are highlighted with a red rhombus and a square, respectively. Data from \citet{Akeson2013}, downloaded on 2021 April 13.}

\label{fig:distrib}
\end{figure}

While many Saturn-sized planets have been found dwelling in extremely close to their host star, the observed distribution of planets shows a depletion of hot Jupiters in orbital periods less than a day (see the grey and green regions in Fig. \ref{fig:distrib}). Also, what was previously a clustering of orbital periods nearing three days, the so-called `three-day pile-up' \citep{Ford2006}, seems to have spread over a wide range of orbital periods excluding, however, those of ultra-short period (USP) hot Jupiters. This shortage of USP hot Jupiters may be the result of stellar tides making them undergo orbital decay and eventually producing their tidal disruption \citep*{Jackson2009}. Such orbital changes should present long-term observational imprints that have only been detected for the WASP-12 system \citep{Maciejewsky2016,Patra2017,Bailey2019,Yee2020}, proving that orbital decay does happen, but leaving the necessity for finding convincing evidence in other favourable candidates like NGTS-10b or WASP-19b: the two shortest-period exoplanets discovered to date.

%%%%%%%%%%%%%%%%%%%%%%%%%%%%%%%%%%%%%%%%%%%%%%%%%%%%%%%%%%%%%%%%%%%%%%%
% FIGURE
%%%%%%%%%%%%%%%%%%%%%%%%%%%%%%%%%%%%%%%%%%%%%%%%%%%%%%%%%%%%%%%%%%%%%%%
%%%%%%%%%%%%%%%%%%%%%%%%%%%%%%%%%%%%%%%%%%%%%%%%%%%%%%%%%%%%%%%%%%%%%%%

Unlike small low-mass planets which are stable against tidal spiral-in \citep{Hamer2020}, intense research is being carried out to gather observational evidence of such a phenomenon in hot Jupiters (see \citealt{Patra2020} and references therein). This, in turn, could help us constrain theories of tidal evolution. Lately, this has been explored in depth by several studies of compact star-planet and planet-moon systems, which have made evident how important is to develop more accurate dynamical models of tidal interactions to contrast theoretical expectations against observed properties and distribution of already discovered systems (see e.g. \citealt*{Dobs2004,Ferraz2008,Jackson2008a,Miller2009,Barker2009}).

Most USP planets have evolved towards tidally-driven circular orbits via tidal torques exerted by the star and the dissipation of the corresponding interaction energy within the planet and the star. Once a circular orbit is reached, further changes in the semi-major axis and planetary spin are produced by two mechanisms: 1) transfer and dissipation of orbital angular momentum to the star and 2) exchange of orbital and planetary angular momenta. The role of both mechanisms in orbital evolution of USP planets makes them valuable for testing models of planetary and stellar interior structure \citep{Brown2011,Penev2016, Penev2018}, as well as for constraining theories of planetary formation and evolution. USP planets are thus very useful 
`probes' for improving our understanding of tidal interactions (see e.g. \citealt{Penev2012, Ginzburg2015}).

Energy dissipation within a body under a forced oscillation is parametrized using two key quantities: 1) the tidal Love coefficient $k_2$, which stands for the {\em level of distortion} that the body can undergo with respect to its equilibrium figure.  This quantity is determined by different physical properties including bulk density and rigidity. And 2) the tidal quality factor $Q$, a dimensionless property that accounts for the fraction of energy dissipated inside the body per tidal forcing cycle (e.g. \citealt{Hansen2010,Zahn2008}). Here, most of our attention will focus on the ratio $k_2/Q$ which corresponds to the imaginary part of the second-order Love number (Section \ref{sec:evolution}). This quantity is directly linked to the tidal dissipation of a star-planet system and affects its rotational and orbital dynamics (Section \ref{sec:tidal}).

The value of these two key quantities, even in the case of well-known objects such as the planets and moons of the Solar System, has been difficult to constrain \citep{Goldreich1966,Yoder1981,Greenberg1982,Greenberg1989,Dickey1994,Lainey2012,Albrecht2012}. Moreover, the evolution of the interior structure of bodies involved in tidal interactions (stars, planets, and moons) changes both quantities in complex ways. Several authors have developed and tested models on which planetary \citep{Jackson2008a,Ferraz2015,Alvarado2019}, and moon semi-major axis (\citealt{Ferraz2008}; \citealt*{Alvarado2017}; \citealt{Sucerquia2019,Sucerquia2020}) evolution have been studied under the coupled evolution of the interior structure and the value of $k_2$ and $Q$. Their results suggest that including the variable nature of these quantities may lead to significant differences in the predicted evolution and fate of interacting systems (e.g. see the different fate of exomoons around close-in planets in \citealt{Alvarado2017}).

Still, we lack a deep understanding of these parameters of exoplanets and stars: what are their typical values? how are they determined by the object properties? and how do they evolve? This is precisely how planets in extremely close orbits could help us disentangle the existing uncertainties in tidal models.

The shortest-period hot Jupiters, WASP-19b \citep{Hebb2010, Wong2016} and  NGTS-10b \citep{McCormac2019}, orbit their host star with a period of less than a day (see Fig. \ref{fig:distrib}) providing a unique opportunity to study the interplay between interior structure, tidal interactions, and orbital decay. Their orbital evolution has been studied using Jupiter-like characteristics for the planets, and typical values of $k_2$ and $Q$ for the stars \citep{Brown2011, Penev2018, McCormac2019}. However, to compute a more accurate tidal-induced evolution we must take into account how the energy dissipation within the planet, and more importantly within the star, is evolving \citep{Alvarado2019}, namely how $k_2$ and $Q$ are dynamically changing with time for both the star and the planet.

In this paper, we aim to reanalyse the orbital decay of these USP planets using an interior structure which adopts a bi-layer composition formalism for the planet \citep{Guenel2014} and the star \citep{Mathis2015b}, along with models relating those properties to the evolution of the key tidal quantities at interplay. As a first step, it is acceptable using a simplified two-layer model to describe the internal structure of the star and the giant planet. However, to constrain $Q$ for giant exoplanets, astronomers often refer to studies of Jupiter \citep{Goldreich1966,Yoder1981,Lainey2009} and Saturn \citep{Lainey2012}. In fact, we now have a deep knowledge of the internal structure and dynamics of these two planets owing to the restless work of the Juno and Cassini space missions (see e.g. \citealt{Kaspi2017,Guillot2018,Galanti2019}), which have also shed light on some of the processes driving their tidal dissipation. Also, other efforts to study the tidal evolution of massive planets have incorporated magnetic braking and tidal history of stars (e.g. \citealt{Bolmont2016}), and studied how the transfer of angular momentum between stellar convective and radiative zones in evolving stars affect the evolution of the whole system \citep{Penev2014, Benbakoura2019}.

This paper is organized as follows. In Section \ref{sec:evolution} we describe the structural model employed and its considerations for planets and stars, whilst in Section \ref{sec:tidal} we present the numerical tidal model with its main assumptions and characteristics. Section \ref{sec:results} contains the implications and results of this approach for the two exoplanets with the shortest orbital period known to date, NGTS-10b and WASP-19b; and Section \ref{sec:dicussion} is devoted to a discussion of orbital decay in an observational context. Finally, the foremost remarks of this work and connections to future research are presented in Section \ref{sec:conclusion}.

%%%%%%%%%%%%%%%%%%%%%%%%%%%%%%%%%%%%%%%%%%%%%%%%%%%%%%%%%%%%%%%%%%%%%%%%%%%%%%%%%%%%%%%%%%%%%%%%%%%%%%%%%%%%%%%%%%%%%%%%%%%%%%%%%%%%%%%%%%%%%%%%%%%%%%%%%%%%
\section{Models for tidal dissipation}
\label{sec:evolution}

For modelling the interior structure of both the star and the planet, we use the formalism of \citet{Remus2012} and \citet{Ogilvie2013}; namely, we assume that the bodies are made only of two different layers (bi-layer model), rotating with the same angular speed (rigid rotation). The latter simplifies the treatment of tidal dissipation as differential rotation impacts tidal gravito-inertial waves \citep*{Ivanov2013, Ogilvie2004} and tidal inertial waves propagating in convective envelopes \citep{Baruteau2013,Favier2014,Guenel2016}. In fact, in a more realistic model, tidal dissipation both in the radiative core and the convective envelope should be included 
\citep{Goodman1998,Terquem1998,Barker2010, Guillot2014,Ogilvie2014,Mathis2019,Barker2020},  although the implications on orbital decay are still debated \citep{Guillot2014}.
 
Body oscillations give rise to complex tidal stress waves which are in general dependent on the tidal frequency $\omega$ (see e.g. \citealt{Remus2012, Efroimsky2012}). The deformation involved on these waves are measured using in general complex coefficients called the Love numbers, $k_l^m(\omega)$. In the most simple case (negligible obliquity of the bodies with respect to the mutual orbital plane), only $k_2^2(\omega)$ is the only non-negligible coefficient.  To study the evolution of the tidal properties of the star and planet, we will use the \textit{Tidal Dissipation Reservoir} (TDR) formalism \citep{Ogilvie2013,Guenel2014}. In this formalism, the frequency-averaged imaginary part of $k_2^2(\omega)$ is used to estimate the ratio $k_2/Q$ of a fluid body (see equation 1 in \citealt{Alvarado2017}) with $\omega\,\epsilon\,[-2\Omega, 2\Omega]$, where $\Omega$ is the rotational rate of the primary body (star or planet).

Adopting a solid-fluid boundary between the two layers in the case of the planet (subscript p) \citep{Guenel2014}, and a fluid-fluid boundary for the star (subscript $\star$) \citep{Mathis2015b}, analytical expressions for the frequency-averaged ratio $\langle k_2/Q\rangle$ can be obtained for characterising the TDR of both the envelope (subscript e) and the core (subscript c). 

In the case of the envelopes we have:

\beq{eq:k2QFormulas}
\begin{split}
    \left<\frac{\Kss}{\Qs}\right>_\mathrm{e} &= \frac{100\pi}{63}\epsilon_\mathrm{\star}^{2}\frac{\alphas^{5}}{1-\alphas^{5}}(1-\gammas^2)(1-\alphas^2)\\&\left(1 + 2\alphas + 3\alphas^2 + \frac{3}{2}\alphas^3\right)^2\left[1+\left(\frac{1-\gammas}{\gammas}\right)\alphas^3\right]\\&\left[1 + \frac{3}{2}\gammas+\frac{5}{2\gammas}\left(1 + \frac{1}{2}\gammas-\frac{3}{2}\gammas^2\right)\alphas^3-\frac{9}{4}(1-\gammas)\alphas^5\right]^{^{-2}}
\end{split}
\eeq
\beq{eq:k2QFormulap}
\left<\frac{\Kpp}{\Qp}\right>_\mathrm{e} = \frac{100\pi}{63}\epsilon_\mathrm{p}^{2}\frac{\alphap^{5}}{1-\alphap^{5}}\left[1+
	\frac{1-\gammap}{\gammap}\alphap^{3}\right]\left[1+
		\frac{5}{2}\frac{1-\gammap}{\gammap}\alphap^{3}\right]^{^{-2}}
\eeq 
where $\epsilon^{2}\equiv (\Omega/\Ocri)^{2}$ is a dimensionless parameter and $\Ocri\equiv(G M/R^{3})^{1/2}$ is the so-called critical rotational rate, defined as the rate at which points on the surface of the body are barely orbiting it; $\Ocri$ is given by Kepler's third law. As $\epsilon\propto\Omega$, Coriolis acceleration is being included in our model \citep{Guenel2014, Mathis2015b}. However, centrifugal forces which scale as $\Omega^2$ are being neglected in the treatment of tidal inertial waves (cf. \citealt{Braviner2014, Braviner2015}). This is an important assumption since tidal interactions on close-in planets, which orbit very fast, often make the planet to be gravitationally locked and high rotational rates can be reached.

The aspect ratios $\alpha$ and $\beta$ quantify the relative size and mass distribution of the body layers:

\beq{eq:k2Qparameters}
\left.\begin{split}
& \alpha \equiv \frac{R_\mathrm{c}}{R}\\ 
& \beta \equiv \frac{M_\mathrm{c}}{M}\\
& \gamma \equiv \frac{\alpha^{3}(1-\beta)}{\beta(1-\alpha^{3})}
\end{split}
\hspace{1.25cm}\right\}
\eeq
with $R$ ($R_\mathrm{c}$) and $M$ ($M_\mathrm{c}$) the total (core) radius and mass of the corresponding body (planet, p, and star, $\star$).

Equations (\ref{eq:k2QFormulas}) and (\ref{eq:k2QFormulap}) represent the contribution to the dissipation reservoir generated by the excitation of inertial waves via viscous friction in convective (fluid) envelopes, assumed as incompressible and homogeneous \citep{Ogilvie2013}. In the case of the star, we assume that this is the only contribution to total tidal dissipation. However, in the case of the planet and to give a more precise estimation of tidal evolution, the tidal dissipation reservoir  $\Kpp/\Qp$ corresponding to the inelastic deformation of a perfectly homogeneous and rigid central core must also be accounted for.

%%%%%%%%%%%%%%%%%%%%%%%%%%%%%%%%%%%%%%%%%%%%%%%%%%%%%%%%%%%%%%%%%%%%%%%
% FIGURE
%%%%%%%%%%%%%%%%%%%%%%%%%%%%%%%%%%%%%%%%%%%%%%%%%%%%%%%%%%%%%%%%%%%%%%%
\begin{figure}
        \hspace{-10pt}
       \includegraphics[scale=0.23]{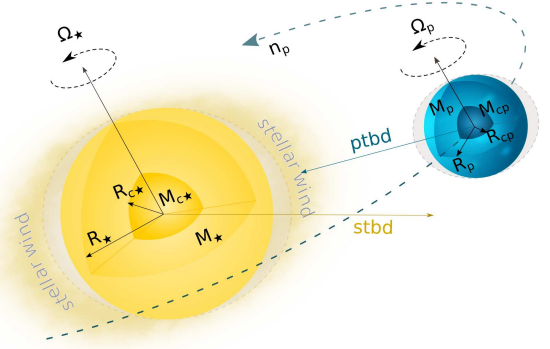}
    \caption{Diagram showing the star and planet along with their main bulk parameters involved in tidal evolution (Section \ref{sec:evolution}). The grey shadowed ellipses represent the equilibrium figures of the star and the planet. The arrows labelled with \textit{stbd} and \textit{ptbd} stand for the stellar and planetary tidal bulge, respectively, and lie on the main ellipsoid tri-axial axes of each body.}
\label{fig:diagram}
\end{figure}
%%%%%%%%%%%%%%%%%%%%%%%%%%%%%%%%%%%%%%%%%%%%%%%%%%%%%%%%%%%%%%%%%%%%%%%

To compute the viscoelastic dissipation of the inner solid core in a planet with bulk rigidity $\mathcal{R}$, we use the generalized Maxwell model for a linear approximation of viscoelasticity. The resulting analytical expression is \citep{Guenel2014}:

\beq{eq:k2Qcore}
\left<\frac{\Kpp}{\Qp}\right>_{\mathrm{c}} = \frac{\upi\mathcal{R}(3 + \mathcal{A})^{2}\mathcal{BC}}{\mathcal{D}(6\mathcal{D} + 4\mathcal{ABCR})}
\eeq
where $\mathcal{A}$, $\mathcal{B}$, $\mathcal{C}$, and $\mathcal{D}$ are auxiliary quantities given by,

\beq{eq:corepars}
\left.\begin{split}
&\mathcal{A} = 1 + \frac{5}{2}\gamma^{-1}\alpha^3(1-\gamma),\\&
\mathcal{B} = \alpha^{-5}(1 - \gamma)^{-2},\\&
\mathcal{C} = \frac{38\upi}{3}\frac{(\alpha\Rp)^4}{G(\beta\Mp)^2},\\&
\mathcal{D} = \frac{2}{3}\mathcal{A}\mathcal{B}(1-\gamma)\left(1+\frac{3}{2}\gamma\right) - \frac{3}{2},
\end{split}
\hspace{0.8cm}\right\}
\eeq
and $G$ is the gravitational constant. It is important to stress that in equation (\ref{eq:k2Qcore}) the rigidity $\mathcal{R}$ should be provided in Pascals (Pa).

Dependency on time of the TDR and hence of the frequency-averaged ratio $\langle k_2/Q\rangle$ arises from two main sources: 1) the planetary and stellar envelope TDR (equations \ref{eq:k2QFormulas} and \ref{eq:k2QFormulap}) which depend on rotational rates $\epsilon_\mathrm{\star}$ and $\epsilon_\mathrm{p}$, and evolve as a result of tidally driven angular momentum exchange (see equations \ref{eq:dosdt} and \ref{eq:dopdt} in Section \ref{sec:tidal}); and 2) the rotation period of the star and its mass change due to stellar magnetic braking and wind-induced mass-loss (see Section \ref{sec:stellarwind}). Under this evolutionary regime, $\beta_\star$ can be written as:

\beq{eq:varybeta}
\betas(t)=\betaso\frac{\Mstaro}{\Mstar(t)}, 
\eeq

\section{Evolutionary models}
\label{sec:tidal} 

Most USP planets have a significantly faster orbital mean motion,  $\npp$, than the rotational rate of their host star (i.e. $\npp\gg\Os$). Whereas these two quantities have been measured for various compact systems, planetary rotational rates, $\Op$, are often unknown. However, considering how close are USP planets from their parent star, we assume that tidal interactions have led them to a synchronous orbital state (i.e. $\npp\sim\Op$) in a time-scale that varies from system to system depending on different physical parameters \citep{Peale1977, Rasio1996}. Also, we neglect any resonances between $\Op$ and $\npp$ (see \citealt{Winn2005}).

To numerically solve all the ordinary differential equations of the next two subsections, we use our own integrator written completely in {\rm Python}, and we dynamically monitor the evolution of the system to adopt a nonstiff-method (Adams) or a stiff-method (with a Backward Differentiation Formula) when necessary. 

The initial orbits of the planets are larger than their critical Roche radius, $\roche$, which is the distance where the tidal stresses of a host star start to overcome the self-gravity of a planet \citep{Roche1849}. The evolution of each star-planet system in this work is integrated until the planet reaches its corresponding $\roche$, defined as

\beq{eq:rtidal}
\roche = \eta\left(\frac{\Mstar}{\Mp}\right)^{1/3}\Rp,
\eeq
where $\eta=2.7$ from simulations by \citet*{Guillochon2011} for the disruption of hot Jupiters.

We proceed now to explain the mathematical treatment of the systems and their evolution, driven by $\npp$, $\Op$, $\Os$, and the planet's orbital eccentricity, $e$.

\subsection{Planetary orbit}
\label{sec:planetorbit}

As we will see later in this work, for different combinations of some parameters, the angular momentum exchange between the planet's orbit and the stellar rotation may produce short or large orbital decay time-scales and slow or fast stellar spin-up rates. Despite the evolution of the systems commences under the given terms explained at the beginning of this Section \ref{sec:tidal}, we will find some cases where, for instance, situations such as $\Os\lesssim\npp$ may arise. Therefore, to work with any possible relations between the stellar rotation and the planet's orbital period, we choose to use a general form of the equations that describe the tidal evolution of the planet's elements (see \citealt{Ferraz2008}). Parallel to \citet{Brown2011}, the instantaneous variation of the planetary mean motion and eccentricity will be given by \citep{Hut1981}:

\beq{eq:dnpdt}
\frac{\dot{n}\sub{p}}{\npp} = -3\left[\frac{\dot{e}}{e}\frac{e^2}{1-e^2}-\frac{\Is\dot{\Omega}_\star}{\Lorb}-\frac{\Ip\dot{\Omega}\sub{p}}{\Lorb}+\frac{\Is\oms}{\Lorb}\right],
\eeq
and,

\beq{eq:dedt}
\begin{split}
\frac{\dot{e}}{e}=\frac{27\npp}{\apos^5}&\Bigg\{\left<\frac{\Kpp}{\Qp}\right>_\mathrm{c+e}\frac{\Mstar(t)}{\Mp(t)}\Rp^5\left[\frac{11}{18}\frac{\Op}{\npp}e_2(e)-e_1(e)\right]\\&+\left<\frac{\Kss}{\Qs}\right>_\mathrm{e}\frac{\Mp(t)}{\Mstar(t)}\Rstar^5\left[\frac{11}{18}\frac{\Os}{\npp}e_2(e)-e_1(e)\right]\Bigg\},
\end{split}
\eeq
where $e_1$ and $e_2$ are functions of $e$ defined as,

\beq{eq:ecc1}
e_1(e) = \left(1 + \frac{15}{4}e^2 + \frac{15}{8}e^4 + \frac{5}{64}e^6\right) \bigg/ (1-e^2)^{13/2},
\eeq

\beq{eq:ecc2}
e_2(e) = \left(1 + \frac{3}{2}e^2 + \frac{1}{8}e^4\right) \bigg/ (1-e^2)^{5},
\eeq
and the orbital angular momentum $\Lorb$ will be

\beq{eq:lorb}
\Lorb = \Mp\Mstar\sqrt{\frac{G\apos(1-e^2)}{\Mstar + \Mp}}
\eeq
In equation (\ref{eq:dnpdt}), $\Is = \eps\gyrs\Mstar\Rstar^2$ and $\Ip = \epp\gyrp\Mp\Rp^2$ are the angular moment of inertia of the star and the planet, respectively. Here, $\gyrs$ ($\gyrp$) is the stellar (planetary) radius of gyration (hereafter \textit{gyradius}), while $\eps$ ($\epp$) stands for the mass fraction in the convective envelope participating in angular momentum exchange. The planet's semi-major axis, $\apos$, will be computed from Kepler's third law as $[G(\Mstar + \Mp)/\npp^2]^{1/3}$. We will assume that the planet's rotational axis is aligned with its orbital plane, and that the star-planet systems in this work are co-planar\footnote{When the orbital plane is inclined, a different framework must be followed, as that studied by  \citet{Barker2009}.}.

It is worth noting that, despite tidal torques act on the whole stellar and planetary convective envelopes, the exchange of angular momentum is a function of the mass of the convective zone. The latter, in turn, is a function of the effective temperature and hence of spectral type \citep*{Pinsonneault2001}. In consonance, the aforementioned dependencies ponder a conundrum: what effect does the mass of the convective zone have on the amount of angular momentum being transferred from the stellar spin to the planetary orbit? To address it, we have followed a similar approach to that of \citet{Dobs2004} and assumed that the planet's dynamical evolution will depend upon the mass fractions $\eps$ and $\epp$. These two quantities were introduced in the expressions for $\Is$ and $\Ip$ of the previous paragraph. Furthermore, the models of Section \ref{sec:evolution} assume that star and planet rotate as single rigid bodies (i.e. without differential rotation), but no preference to where the tidal dissipation occurs was made. However, if most of the angular momentum comes from the convective zones, any disparity in their mass may result in an over- or under-estimation of $\koQs$ (i.e. tidal dissipation) affecting thus the orbital evolution of USP planets.

\subsection{Rotational rates}
\label{sec:rotationrates}

For this work, rotational angular momentum in both the star and the planet will only be produced from the outer convective envelope. In other words, we neglect any transfer of angular momentum made by the stellar/planetary inner solid core. For a star aligned with the planet's orbital plane, in a system where $\Mstar\gg\Mp$, the evolution of the rotational rates $\Os$ and $\Op$ is given by \citep{Mardling2002,Dobs2004},

\beq{eq:dosdt}
\frac{\Der\Os}{\Der t}= \frac{3\npp^4\Mp(t)^2}{\eps\gyrs G}\left(\frac{\Rstar}{\Mstar}\right)^3\left<\frac{\Kss}{\Qs}\right>_\mathrm{e}\left[e_3(e)-e_4(e)\left(\frac{\Os}{\npp}\right)\right] + \oms,
\eeq
and

\beq{eq:dopdt}
\frac{\Der\Op}{\Der t}= \frac{3\npp^4\Rp(t)^3}{\epp\gyrp G\Mp(t)}\left<\frac{\Kpp}{\Qp}\right>_\mathrm{c+e}\left[e_3(e)-e_4(e)\left(\frac{\Op}{\npp}\right)\right] + \omp,
\eeq
where $e_3$ and $e_4$ are functions of $e$ defined as,

\beq{eq:ecc3}
e_3(e) = \left(1 + \frac{15}{2}e^2 + \frac{45}{8}e^4 + \frac{5}{16}e^6\right) \bigg/ (1-e^2)^6,
\eeq

\beq{eq:ecc4}
e_4(e) = \left(1 + 3e^2 + \frac{3}{8}e^4\right) \bigg/ (1-e^2)^{9/2}.
\eeq
In equations (\ref{eq:dosdt}) and (\ref{eq:dopdt}), $\oms$ ($\omp$) are the angular momentum braking rate due to stellar and planetary winds. Similar to $\koQs$ and $\koQp$, $\eps$ ($\epp$) are also unknown quantities which do not have a stringently or widely accepted formulation, yet the tidal evolution of USP planets can strongly depend on them, as described in Section \ref{sec:results}. Nevertheless, at least for planets, we can make a well-educated assumption \citep{Dobs2004}: since convection zones in gaseous planets are significantly large they are assumed to be fully mixed, so $\gyrp\epp\simeq\gyrp$ (i.e. $\epp\simeq1$). On the contrary, the convection zones in stars are shallower, and both $\gyrs$ and $\eps$ are unknown functions of the stellar spectral type. Therefore, our study will encompass several values for the product $\gyrs\eps$, from low- to high-mass convective envelopes.

\subsection{Stellar and planetary mass}
\label{sec:stellarwind}

To use a more robust self-contained tidal model, the planet's mass-loss during orbital decay will also be accounted for. To that end, the mass of the planet will change as follows,

\beq{eq:Mloss-rate}
\Mprate = -\frac{\pi \Rp^{3} F_\mathrm{XUV}}{GK\Mp(t)}- \left(\frac{\Rp(t)}{\apos}\right)^2 \frac{\Mstarrate \alpha}{2}.
\eeq
with $\Mstarrate$ the mass loss rate of the star due to the stellar wind, $K$ a parameter representing the planet radius losses at the Roche lobe, $F_\mathrm{XUV}$ the stellar flux in X-rays and Extreme Ultra Violet at the planet's position, $G$ the gravitational constant, and $\alpha=0.3$ an entrainment efficiency factor. In equation (\ref{eq:Mloss-rate}), which includes the mass-loss due to atmospheric photo-evaporation (first term) and stellar wind drag (second term), all quantities will be set following \citet*{Zendejas2010} and \citet{Sanz-Forcada2011}, excepting $\Mstarrate$ which we assume as a power-law function that depends on the stellar mass and rotation as follows,

\beq{eq:stellarmassrate}
\Mstarrate =\left(\frac{\Mstar}{\Msun}\right)^{a}
\left(\frac{\Os}{\Osun}\right)^{b} \left(\frac{\Rstar}{\Rsun}\right)^{2}\Msunrate
\eeq
where $a=-3.36$, $b=1.33$, $\Osun=2.67\times10^{-6}$ rad s$^{-1}$, and $\Msunrate=1.4\times10^{-14}\Msun$ yr$^{-1}$ \citep{Johnstone2015}.

Also, as the planet's mass is changing via equation (\ref{eq:Mloss-rate}), the planetary radius should change accordingly. However, it is not well-settled how the radius of giant planets vary with their mass. In fact, there could not be an analytic and straightforward method for modelling such size-related variations in USP planets. Therefore, and for the sake of consistency and completeness in the tidal model used in this work, we will adopt $\Rp=c\Mp
^{0.01\pm0.02}$ from \citet{Bashi2017} (cf. \citealt{Chen2017}), where the constant $c$ will be set using measured initial values of mass ($\Mpo$) and radius ($\Rpo$).

The main cause for the gain of rotational angular momentum in the star is the transfer of orbital angular momentum from its USP planet. At the same time, USP planets are also affected by the stellar wind (see equation \ref{eq:dopdt}), which delays their orbital decay. Still, in post-formation scenarios, it can be assumed that $\omp\simeq0$, and these are denominated stable and conservative systems \citep{Dobs2004}. Also, the star itself spins down due to the loss of angular momentum via the stellar wind (i.e. $\oms$ in equation \ref{eq:dosdt}), and for a more robust calculation of a system's evolution, we should account for this effect which might be computed using the average stellar rotational velocity according to \citet{Skumanich1972}. By following \citet{Weber1967}, we compute $\oms$ as follows,

\beq{eq:magbrak}
\oms=-\kappa \Os \mathrm{Min}(\Os, \Obar)^2
\eeq
where $\kappa$ is a constant of proportionality that determines the physical scaling of the magnetic braking, and $\Obar$ is the so-called `saturation' rate, the upper limit where the relationship of the magnetic dynamo and the stellar spin breaks down, also known as the saturation regime. $\Obar$ is an unknown function that depends mainly on the stellar spectral type, whose value is highly uncertain and needs empirical estimations. For the model used in this work, we will adopt those empirical values given in \citet{Cameron1994} for stellar masses ranging from 0.7 to 1.0 $\Msun$, as both NGTS-10b and WASP-19b belong to this range.

From the standard model of \citet{Weber1967}, the rate at which the stellar rotation changes due solely to the stellar wind is described by $\Dot{\Omega}\sub{\star}=-\kappa\Os^3$. If we integrate,

\beq{eq:integrationwind}
\int_{\Omega\sub{\star, 1}}^{\Omega\sub{\star, 2}}\Os^{-3}\Der\Os = \int_{t_1}^{t_2}dt
\eeq
assuming that $\Omega\sub{\star, 2}\ll\Omega\sub{\star, 1}$ and $t_2\gg t_1$, we will take $\Omega\sub{\star, 2}\approx\Osini\approx\Os$ and $t_2\approx \tage$, where $\Osini$ is the stellar rotational rate at the initial rotation period (i.e. just before the planet undergoes orbital decay), and $\tage$ is the age of the star. Under these two considerations, $\kappa$ will be given by

\beq{eq:kappa}
\kappa = \frac{\Os^{-2}}{2\tage}.
\eeq
Since the braking torque has been shown to depend on the stellar spin rate \citep{Matt2008, Matt2012, Reville2015a, Reville2016a, Finley2017}, with equation (\ref{eq:kappa}) we allow $\kappa$ to evolve with $\Os$.

\section{Results}
\label{sec:results}
In this section, we present the results of performing a set of numerical simulations for the two shortest-period planets discovered so far, namely NGTS-10b and WASP-19b (see Fig. \ref{fig:distrib}). We found that using different values for the orbital and physical parameters of both planets within their uncertainty ranges produced a negligible difference in the results, so we decided to use the central value only. These two planets were chosen because their `quick' orbital decay might produce a detectable short-term advance in their photometric mid-transit times that could be measured in the next decade \citep{Cameron2018}. This would prove that we are indeed measuring in real-time the shrinking of exoplanetary orbits, and thus we could improve our theories and mathematical treatment when studying tidal interactions.

\begin{figure}
    \includegraphics[scale=0.78]{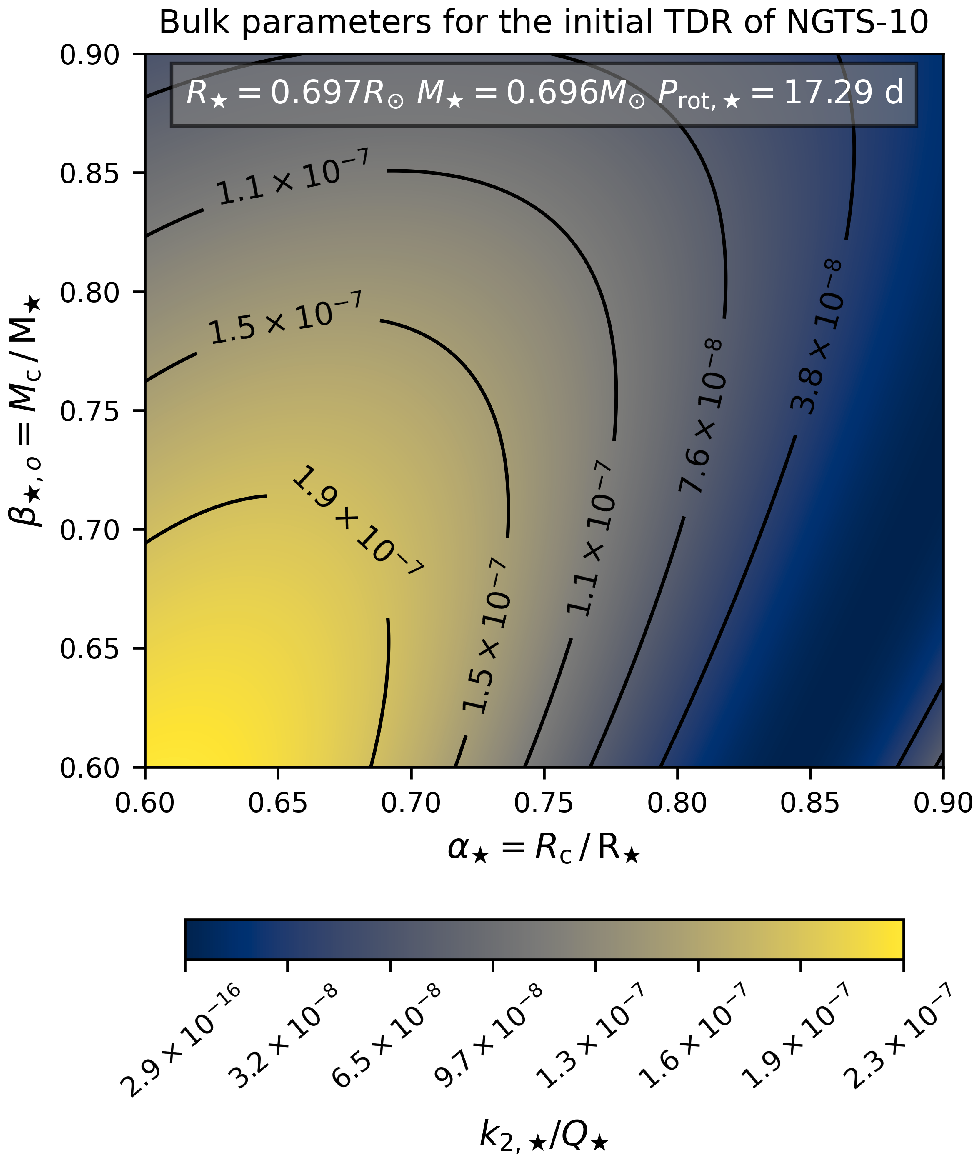}
    \includegraphics[scale=0.78]{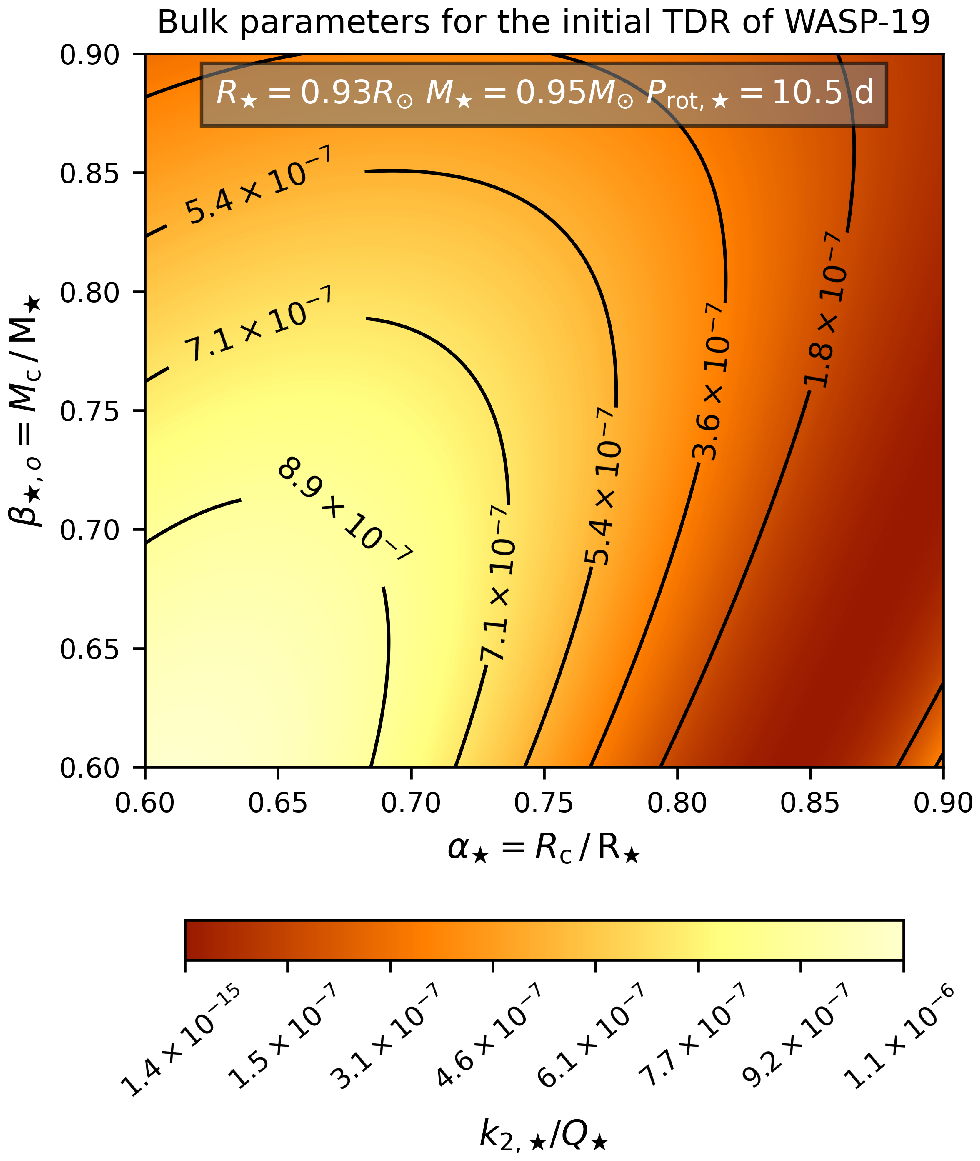}
    \caption{Contour map of bulk parameters for the TDR of NGTS-10 (top panel) and WASP-19 (bottom panel).\label{fig:contours}}
\end{figure}

The mass and size of the core of WASP-19 and NGST-10 are certainly unknown quantities. Therefore, to choose fiducial values for the corresponding quantities in our model, we plot in Fig. \ref{fig:contours} contour maps of $\koQs$ corresponding to different values of $\alpha_\star$ and $\beta_{\star,0}$, for NGTS-10 (top panel) and WASP-19 (bottom panel).

Both quantities, $\alphas$ and $\betaso$ are varied between 0.6 and 0.9 to cover the $\alpha-\beta$ regions of stars from 0.6 to 1.0 $\Msun$ \citep{Gallet2017}. The resulting $\koQs$ lies between $2.9\times10
^{-16}-2.3\times10^{-7}$ for NGTS-10 and $1.4\times10^{-15}-1.1\times10^{-6}$ for WASP-19. Within these ranges, we constrain our model to nominal values of $\Kss/\Qs$  values based on relationships of the stellar TDR and tidal oscillation periods given in terms of the planetary orbital period, $\porb$, and the stellar rotation period, $\prot$, as explained in \citet{Penev2018}.

Despite several parameters being involved in the orbital decay of USP planets, for this work we especially explore the importance that the stellar core's mass (radius) fraction, $\alphas$ ($\betas$), and the product of the stellar inertial momentum and envelope mass, $\eps\gyrs$, have on the tidal interactions of planets and their host star. To constrain the $\alphas-\betas$ parameter space when analysing orbital decay, we used Fig. \ref{fig:contours} to keep $\koQs$ within accepted values computed from stellar tidal spin-up \citep{Penev2018}, so that $\koQs$ ranged from $\sim10^{-8}-10^{-6}$, where small/large values correspond to less/more efficient dissipation.

%%%%%%%%%%%%%%%%%%%%%%%%%%%%%%%%%%%%%%%%%%%%%%%%%%%%%%%%%%%%%%%%%%%%%%
% FIGURE
%%%%%%%%%%%%%%%%%%%%%%%%%%%%%%%%%%%%%%%%%%%%%%%%%%%%%%%%%%%%%%%%%%%%%%
%%%%%%%%%%%%%%%%%%%%%%%%%%%%%%%%%%%%%%%%%%%%%%%%%%%%%%%%%%%%%%%%%%%%%%%

% FIGURE
%%%%%%%%%%%%%%%%%%%%%%%%%%%%%%%%%%%%%%%%%%%%%%%%%%%%%%%%%%%%%%%%%%%%%%%
\begin{figure*}
 \includegraphics[scale=1.1]{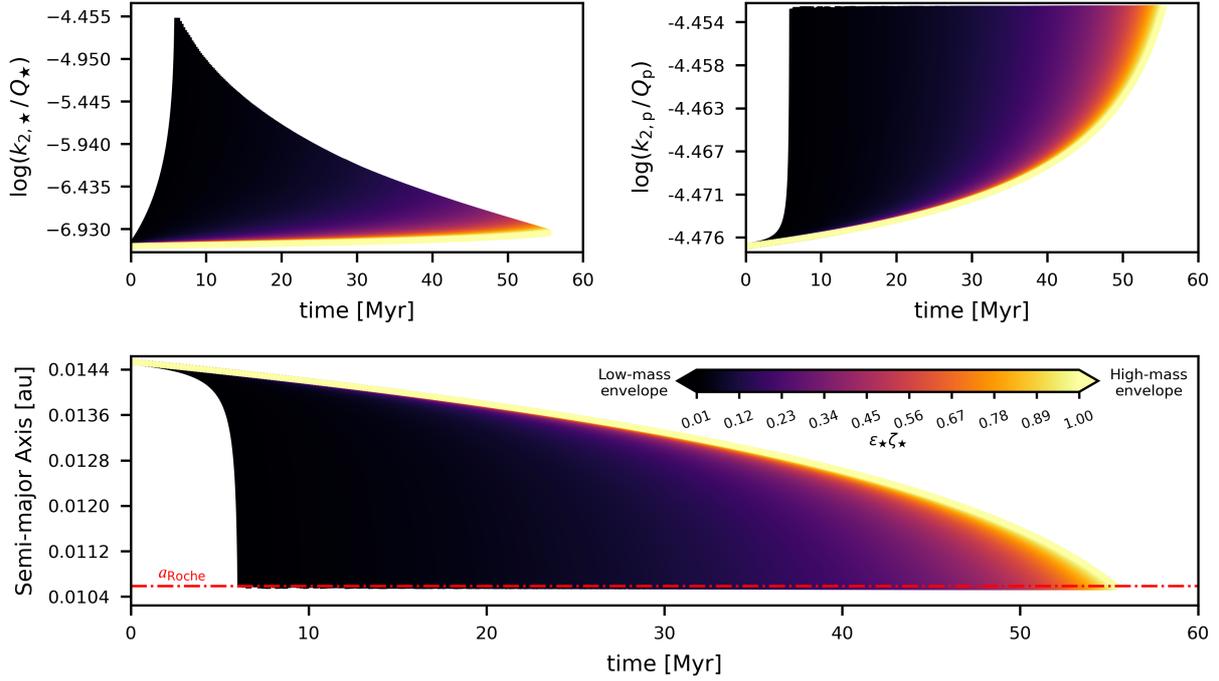}
    \caption{Top panel: evolution of the stellar (left) and planetary (right) TDR. Bottom panel: planetary semi-major axis as a function of time. Plots in this Figure are for NGTS-10b, where the colour map represents different products of the stellar gyradius ($\gyrs$) and the star's fluid envelope mass fraction ($\eps$), ranging from 0.01 to 1.0.}
\label{fig:resultNGTS}
\end{figure*}
%%%%%%%%%%%%%%%%%%%%%%%%%%%%%%%%%%%%%%%%%%%%%%%%%%%%%%%%%%%%%%%%%%%%%%%

%%%%%%%%%%%%%%%%%%%%%%%%%%%%%%%%%%%%%%%%%%%%%%%%%%%%%%%%%%%%%%%%%%%%%%
% FIGURE
%%%%%%%%%%%%%%%%%%%%%%%%%%%%%%%%%%%%%%%%%%%%%%%%%%%%%%%%%%%%%%%%%%%%%%
\begin{figure}
        \includegraphics[scale=0.6]{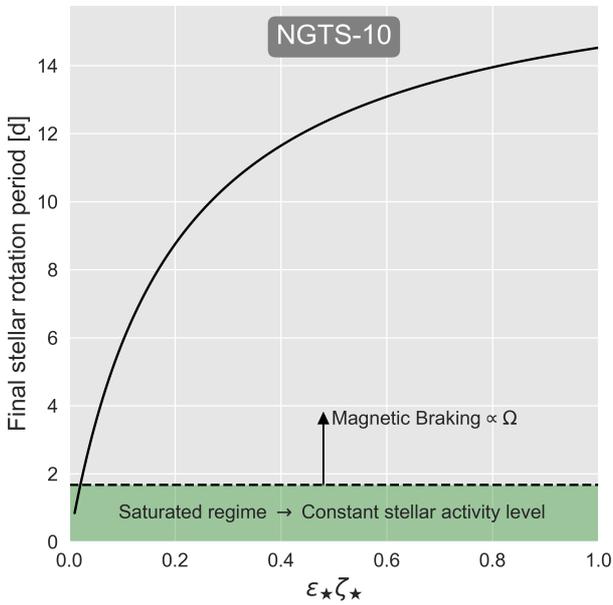}
    \caption{Final stellar rotation period for NGTS-10 as a function of the product of the stellar gyradius ($\gyrs$) and the star's fluid envelope mass fraction ($\eps$).}
\label{fig:finalperiodNGTS}
\end{figure}
%%%%%%%%%%%%%%%%%%%%%%%%%%%%%%%%%%%%%%%%%%%%%%%%%%%%%%%%%%%%%%%%%%%%%%%

Since the planet is losing mass as indicated by equation (\ref{eq:Mloss-rate}), through the physical processes explained in Section \ref{sec:stellarwind}, $\alphap$ and $\betap$ in equation (\ref{eq:k2Qparameters}) will be functions of time, with their instantaneous changes given by $\alphap(t)=\alphapo\Rpo/\Rp(t)$ and $\betap(t)=\betapo\Mpo/\Mp(t)$, respectively. The initial bulk interior properties were taken from the nominal fractions of Jupiter's core mass ($\betapo=0.02$) and radius ($\alphapo=0.126$), which we assume parallel to those of close-in hot Jupiters. Such values are extracted from \cite{Mathis2015a}, who calculated the aforementioned properties for the giant planets in our Solar System (see table 1 in that same work).

%%%%%%%%%%%%%%%%%%%%%%%%%%%%%%%%%%%%%%%%%%%%%%%%%%%%%%%%%%%%%%%%%%%%%%
% FIGURE
%%%%%%%%%%%%%%%%%%%%%%%%%%%%%%%%%%%%%%%%%%%%%%%%%%%%%%%%%%%%%%%%%%%%%%
\begin{figure*}
        \hspace{-12pt}
        \includegraphics[scale=0.44]{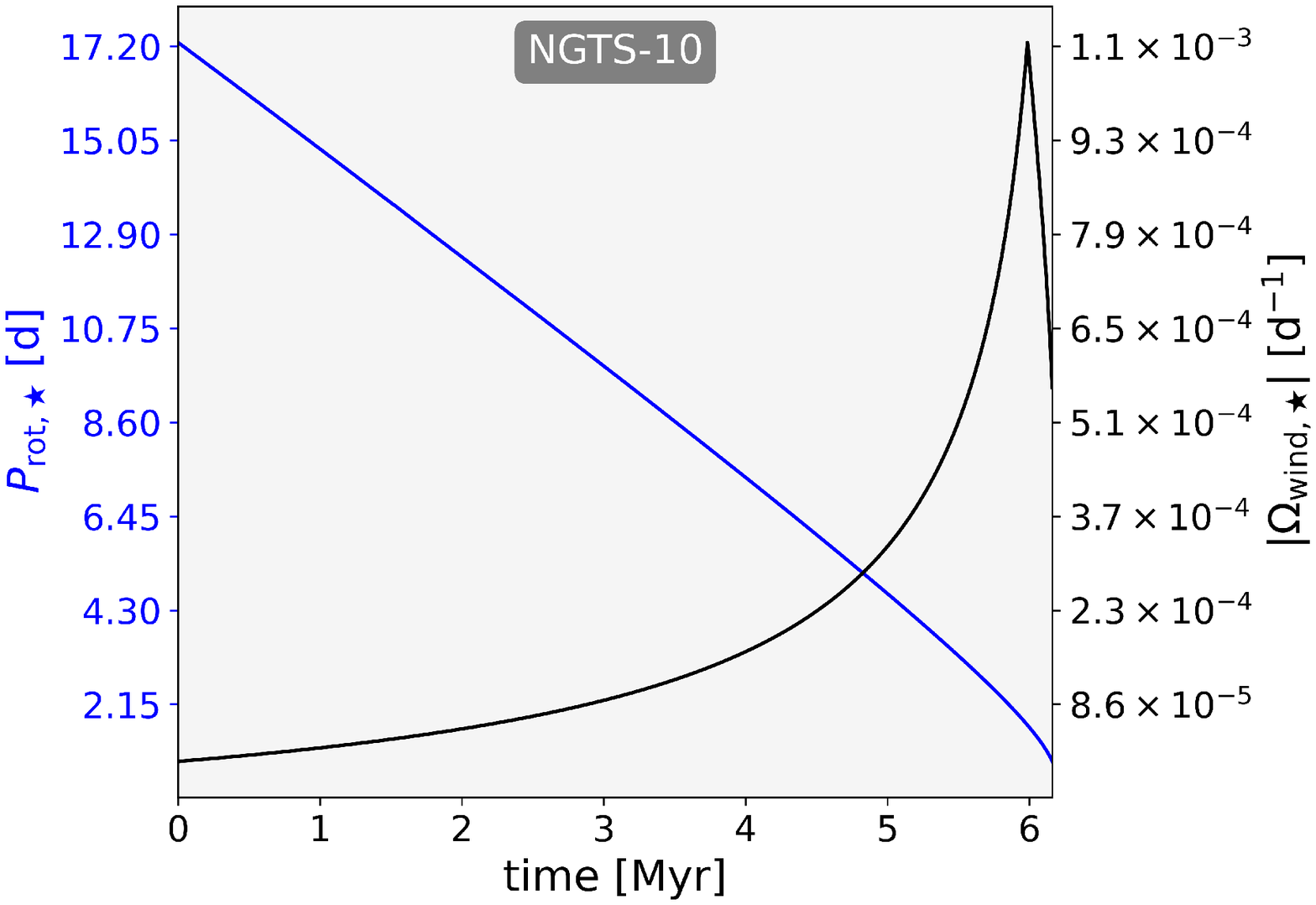}
        \includegraphics[scale=0.44]{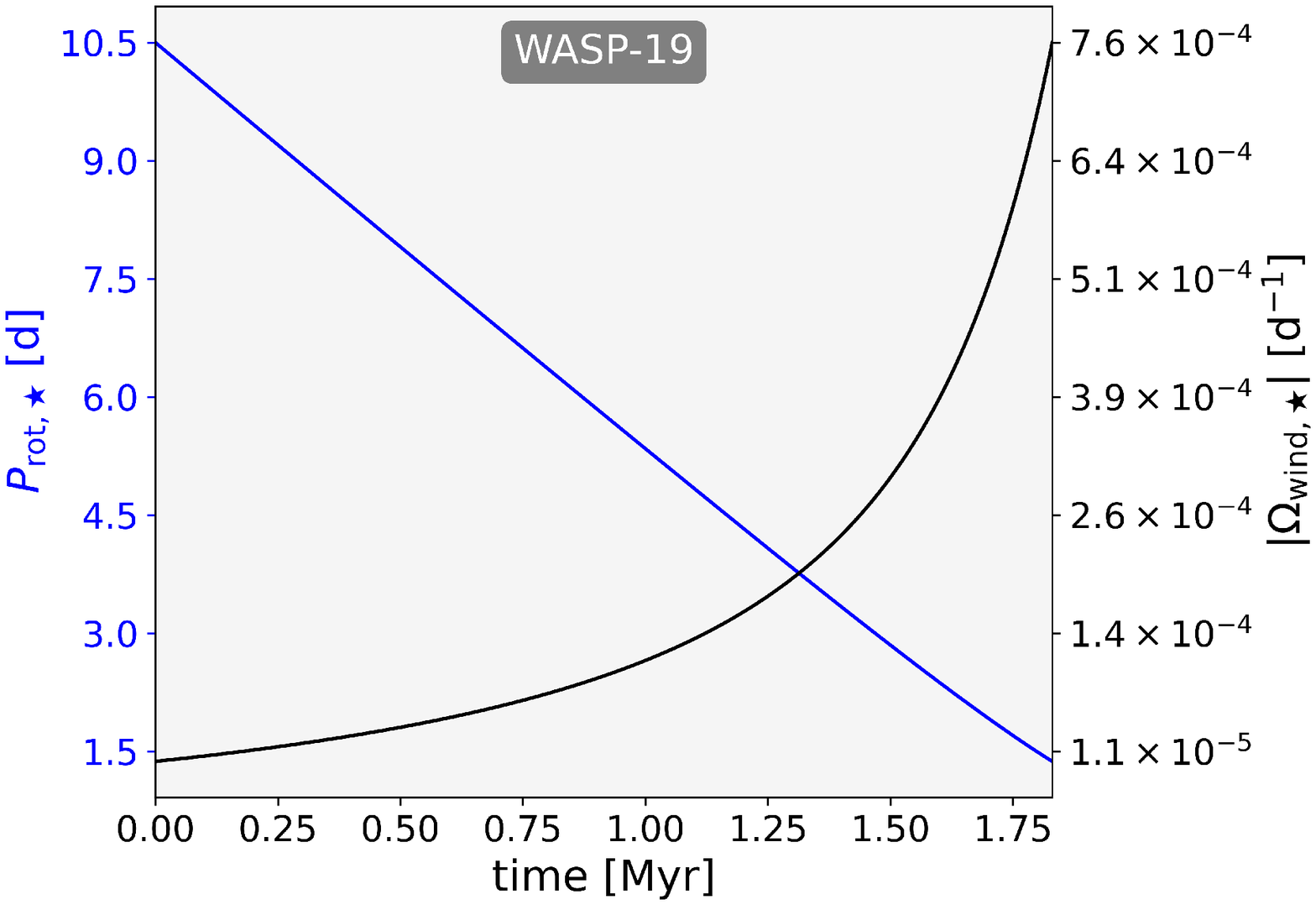}
    \caption{Left-hand panel: evolution of the stellar rotation period (i.e. blue line and y-axis) and the stellar wind rate (i.e. black line and y-axis) for NGTS-10 using $\eps\gyrs=0.01$. Right-hand panel: same as the left-hand panel but for WASP-19 and $\eps\gyrs=0.001$. In both systems, the spin-up of the star produced by the in-falling planet triggers more activity of the stellar wind.}
\label{fig:omegaspin}
\end{figure*}
%%%%%%%%%%%%%%%%%%%%%%%%%%%%%%%%%%%%%%%%%%%%%%%%%%%%%%%%%%%%%%%%%%%%%%%

\subsection{The exoplanet NGTS-10b}
\label{sec:ngts10b}

For the stellar and planetary physical/orbital parameters of the NGTS-10 system, we use the reported values by \citet{McCormac2019}. NGTS-10b is a USP planet with $\porb=0.7669$ d, $\Rp=1.205\, \Rjup$, and $\Mp=2.162\, \Mjup$. Practically speaking, the orbit of this planet is fairly circular, so we have set the eccentricity $e_\mathrm{o}=1\times10^{-6}$ to avoid numerical problems when integrating equations presented in the previous section. The host star of this system ($\Mstar=0.696\, \Msun$, $\Rstar=0.697\, \Rsun$, and $\prot=17.290$ d) is a K-type main-sequence star with an age of $10.4\pm2.5$ Gyr. For NGTS-10 we adopted $\alphas=0.65$ and $\betaso=0.9$ from the top panel of Fig. \ref{fig:contours} based on \citet{Gallet2017}.

We have analysed planetary tidal evolution under the evolutionary mechanism proposed and described in the previous sections, which is worthy of attention when seeking a precise measurement of orbital decay time-scales. For studying the tidal evolution of NGTS-10b both the stellar and planetary TDR (i.e. $k_2/Q$) are evolving, but only the changes in $\Kss/\Qs$ are significant to produce any considerable shift from previously computed orbital decay time-scales.

A notable response to the shrink of NGTS-10b's orbit is the spin-up of its host star due to the transfer of orbital angular momentum. However, the evolution of the stellar rotation has two main contributions, namely that produced by the planet's spiral-in (i.e. orbital decay) and the spin-down produced by the stellar wind; thus, it was worth studying thoroughly these phenomena to fully understand how the NGTS-10 system evolved to its final state. 

From Fig. \ref{fig:omegaspin} we can see how the spin-up of stars (blue line) produced by in-falling planets triggers a more significant stellar wind rate (black line). In the case of NGTS-10 depicted in the left-hand panel of Fig. \ref{fig:omegaspin}, we can see that the angular momentum carried away by magnetic braking increases one order of magnitude, and at the end of the planet's evolution (upon the arrival to the Roche limit), the stellar saturated regime was reached (i.e. the peak of the black line). For WASP-19 (right-hand panel of Fig. \ref{fig:omegaspin}), the spin-up triggered by WASP-19b produces a change of two orders of magnitude in the stellar wind rate that might start driving the transfer of angular momentum. However, despite the change in the stellar wind rate was highly significant, WASP-19 did not reach the saturated regime for any value of $\eps\gyrs$ as can be seen in Fig. \ref{fig:finalperiodWASP}.

For the stellar parameters $\eps\gyrs$ in the left-hand panel of Fig. \ref{fig:omegaspin}, we adopted the leftmost value of the colour bar in Fig. \ref{fig:resultNGTS}. This means that for the analysis made in the previous paragraph, we chose the most extreme value for a low-mass convective envelope of a K5V star (NGTS-10). Given that $\eps$ and $\gyrs$ are highly unknown parameters for most stars, as long as there is no information about their interior structure, that is, to know exactly what the distribution of mass inside those stars is, the only way to study the effect of $\eps$ and $\gyrs$ is to make educated assumptions using what we know about their spectral types.

In light of the above, since $\eps$ and $\gyrs$ appear together in the equation of $\frac{\Der\Os}{\Der t}$ (see Section \ref{sec:rotationrates}), we studied the evolution of the system for different $\eps\gyrs$ within a range of feasible values for a K5V star (i.e. $0.01\leq\eps\gyrs\leq1.0$). As can be noticed from the bottom panel of Fig. \ref{fig:resultNGTS}, different $\eps\gyrs$ lead to a distinct tidal evolution of the planet, and the main effect of varying $\eps\gyrs$ is reflected on the time-scale at which planetary orbital decay towards the Roche radius occurs. Interestingly, when using $\koQs=7.5\times10
^{-8}$, \citet{McCormac2019} reported a median \textit{inspiral} time of 38 Myr assuming $\eps=1.0$ \citep{Brown2011}. However, we can see from Fig. \ref{fig:resultNGTS} that given a value of $\koQs$, orbital decay time-scales may vary over one order of magnitude, meaning that not only the dissipation of tidal energy is important for studying the orbital evolution of NGTS-10b resulting from the interaction of $\Os$ with $\npp$, but also the fraction of stellar mass in the convective fluid envelope is an important part in the exchange of angular momentum.

The largest orbital decay time-scales correspond to NGTS-10 having a high-mass envelope which would produce high rotational inertia. This hinders the angular momentum transfer from the planet's orbit, slowing down the planetary orbital decay process and the spin-up of the star, which means large final stellar rotation periods (see Fig. \ref{fig:finalperiodNGTS}). On the contrary, for a low-mass convective envelope, the star goes quickly from a low- to a high-efficiency dissipation rate of tidal energy, producing a momentary stellar spin-down. However, as the star does not have significant inertia (i.e. small values of $\eps\gyrs$), the decaying orbit of the planet rapidly transfers orbital angular momentum so that it accelerates the stellar rotation, enhancing thereby the energy dissipation within the star.

Also, as depicted on the top-left panel of Fig. \ref{fig:resultNGTS}, depending on the selected value of $\eps\gyrs$ there might be a fast or slow evolution for $\koQs$, which is directly coupled to the transfer of orbital angular momentum and, hence, to the spin-down or spin-up of the star. On the other hand, the evolution of planetary tides (see the top-right panel of Fig. \ref{fig:resultNGTS}) does not have a significant effect on the overall tidal evolution of the planetary orbit. This is due mainly to the solid and inelastic nature of the core's TDR, which is the dominant part (equation \ref{eq:k2Qcore}) and does not change with $\Op$ as the envelope does (equation \ref{eq:k2QFormulap}).

Energy dissipation in the planet's inelastic core was included to better understand the significance of planetary tides in the orbital decay of USP planets. Additionally, we accounted for the mass-loss of their fluid envelope via photo-evaporation and stellar wind drag and found that it would be their nuclei which lead most of the tidal dissipation involved in orbital decay. That said, the planetary core properties upon which $\koQp$ depends may evolve through $\Rp^{-1}$ and $\Mp^{-1}$ (see $\alphap$ and $\betap$ in equation \ref{eq:k2Qparameters}). Still, we found that orbital decay time-scales for NGTS-10b (and later WASP-19b) were too short in comparison to their physical changes and the planetary tidal dissipation remains nearly static throughout the evolution of the system. For the solid core of NGTS-10b, we have assumed a Jupiter-like rigidity (i.e. $\mathcal{R}=4.46\times10
^{10}$ Pa; see \citealt{Guenel2014}).

Finally, a thorough analysis of how the final rotation period of NGTS-10 depends on the stellar angular momentum inertia is given in Fig. \ref{fig:finalperiodNGTS}. In this Figure, we can see that the smaller the rotational momentum inertia the faster the star will rotate at the end of the planet's orbital evolution, whereas large values of $\eps\gyrs$ produce a much slower final stellar rotational rate. For most of the explored values of $\eps\gyrs$ in Fig. \ref{fig:finalperiodNGTS}, NGTS-10 ended with final rotation periods in the regime where magnetic braking is proportional to the stellar rotational rate. For $\eps\gyrs\lesssim0.05$, however, NGTS-10 evolved towards final rotation periods which are within the saturated regime (represented by the green area in Fig. \ref{fig:finalperiodNGTS}) where the stellar activity is fairly constant and scenarios such as $\Os\lesssim\npp$ may arise. Furthermore, we found that regardless of the adopted $\eps\gyrs$ within the feasible range of values of a K-type star, NGTS-10 spun up considerably with respect to its initial rotation period, having final rotation periods between 1.0 d$\leq\prot\leq14.2$ d.

\subsection{The exoplanet WASP-19b}
\label{sec:wasp19b}

The stellar and planetary physical/orbital parameters of the WASP-19 system are taken from the $e$-free MCMC analysis by \citet{Hebb2010}. WASP-19b is a USP planet with $\porb=0.78884$ d, $\Rp=1.28\, \Rjup$, and $\Mp=1.14\, \Mjup$. In this case, the orbit has not been damped to extremely low values of the eccentricity, which is set to $e_\mathrm{o}=0.02$. The host star of this system ($\Mstar=0.95\, \Msun$, $\Rstar=0.93\, \Rsun$, and $\prot=10.5$ d) is a G-type main-sequence star. As discussed in \citet{Brown2011}, the age of WASP-19 is still under constant scrutiny, so we chose a conservative value from an isochrone fitting which puts a lower bound on its age ($\tage\ga1$ Gyr), and WASP-19 was assumed to be $5.5_{-4.5}^{+9.0}$ Gyr old, in concordance with \citet{Hebb2010}.

% FIGURE
%%%%%%%%%%%%%%%%%%%%%%%%%%%%%%%%%%%%%%%%%%%%%%%%%%%%%%%%%%%%%%%%%%%%%%%
\begin{figure}
\hspace{-15pt}
 \includegraphics[scale=0.66]{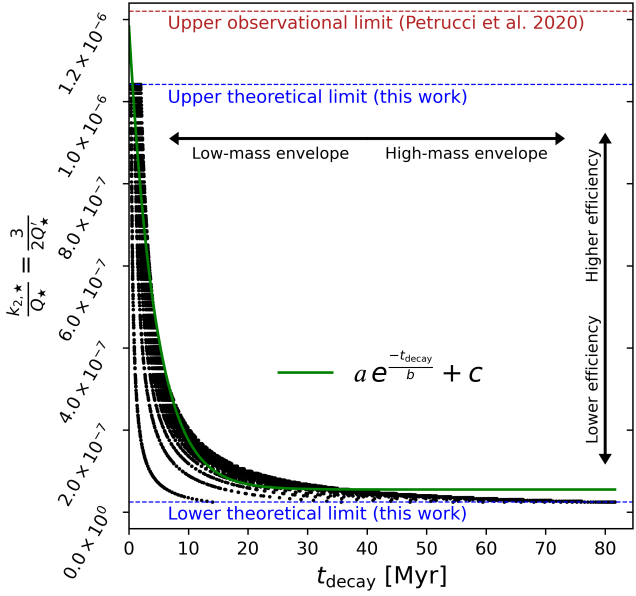}
    \caption{Tidal Dissipation Reservoir of WASP-19 as a function of the orbital decay time-scale of WASP-19b. The horizontal red and blue line represents the observational \citep{Petrucci2020} and theoretical (this work) upper limit for $\koQs$, respectively. See the main text for a complete description of the best-fitting function (green line) and its parameters $a$, $b$, and $c$.}
\label{fig:WASPk2qdecay}
\end{figure}
%%%%%%%%%%%%%%%%%%%%%%%%%%%%%%%%%%%%%%%%%%%%%%%%%%%%%%%%%%%%%%%%%%%%%%%

% FIGURE
%%%%%%%%%%%%%%%%%%%%%%%%%%%%%%%%%%%%%%%%%%%%%%%%%%%%%%%%%%%%%%%%%%%%%%%
\begin{figure*}
        \includegraphics[scale=1.1]{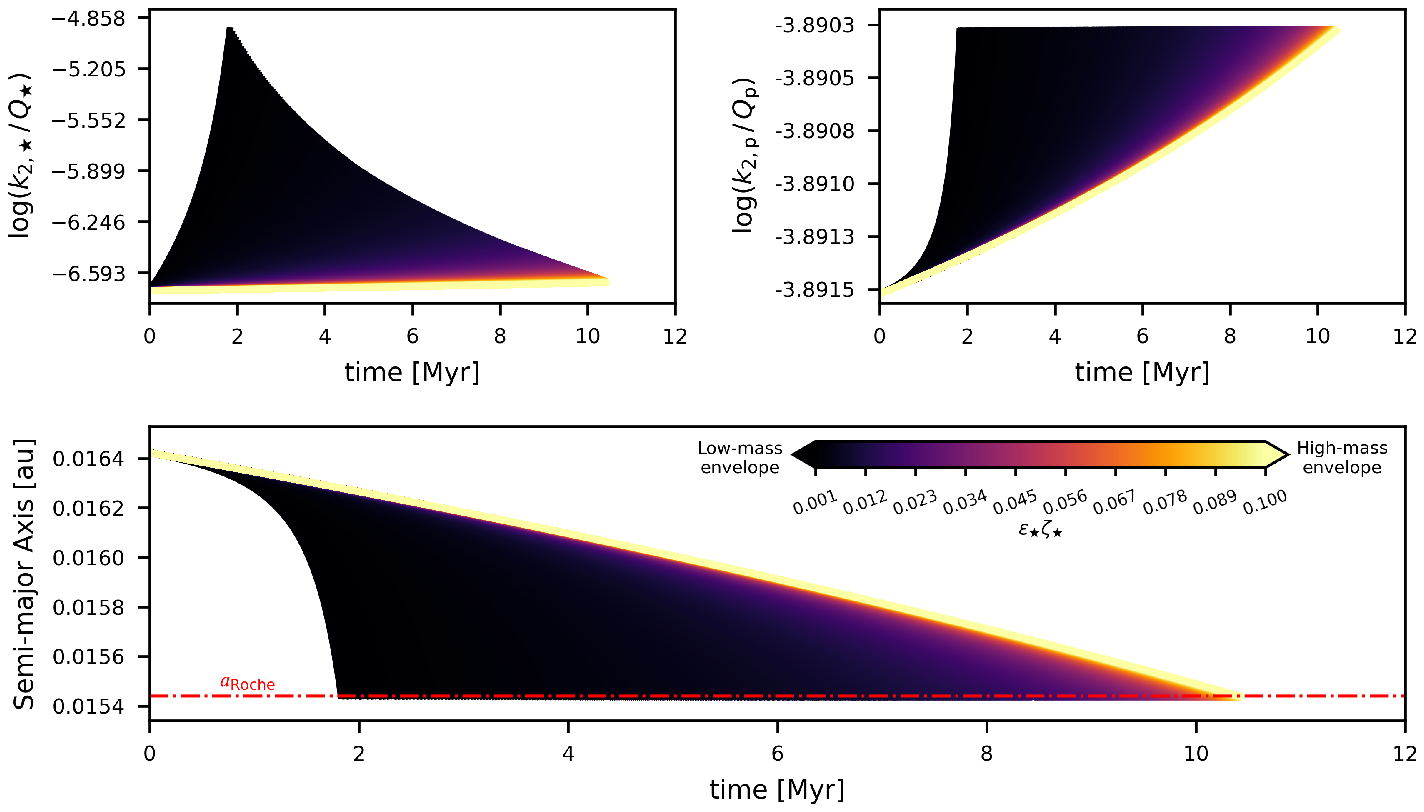}
    \caption{Same as Fig. \ref{fig:resultNGTS} but for WASP-19b and $\eps\gyrs$ ranging from 0.001 to 0.1.}
\label{fig:resultWASP}
\end{figure*}
%%%%%%%%%%%%%%%%%%%%%%%%%%%%%%%%%%%%%%%%%%%%%%%%%%%%%%%%%%%%%%%%%%%%%%%

%%%%%%%%%%%%%%%%%%%%%%%%%%%%%%%%%%%%%%%%%%%%%%%%%%%%%%%%%%%%%%%%%%%%%%
% FIGURE
%%%%%%%%%%%%%%%%%%%%%%%%%%%%%%%%%%%%%%%%%%%%%%%%%%%%%%%%%%%%%%%%%%%%%%
\begin{figure}
         \includegraphics[scale=0.6]{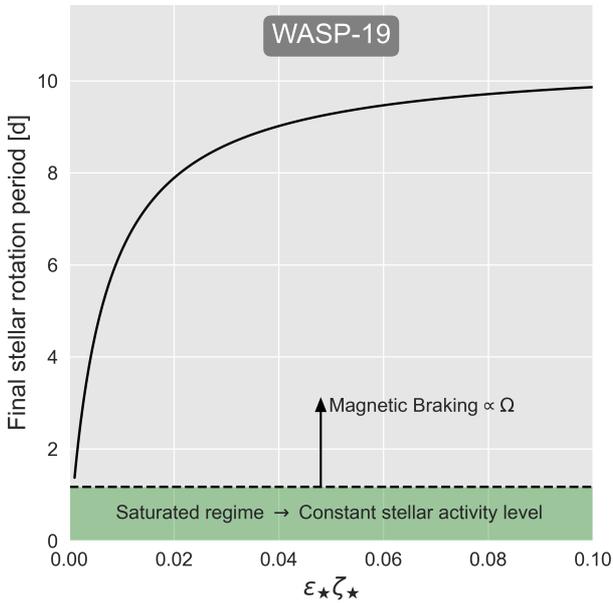}
    \caption{Same as Fig. \ref{fig:finalperiodNGTS} but for WASP-19.}
\label{fig:finalperiodWASP}
\end{figure}
%%%%%%%%%%%%%%%%%%%%%%%%%%%%%%%%%%%%%%%%%%%%%%%%%%%%%%%%%%%%%%%%%%%%%%%

In line with previous works (e.g. \citealt{Hebb2010, Brown2011}), we found that large values of $\koQs$ would produce a notably fast orbital decay. However, as reported by \citet{Petrucci2020}, this would not be the case and there would exist an upper bound for the tidal dissipation of WASP-19 at $\koQs=1.22\times10^{-6}$, represented by the dashed red line in Fig. \ref{fig:WASPk2qdecay}. From this same Figure, we can notice that the tidal evolution model used here allowed us to find an upper theoretical limit of $\koQs=1.042\times10^{-6}$, almost matching the observational constraint of \citet{Petrucci2020}. We found that if $\koQs$ is close to either limit the planet would not survive for longer than $t\sub{decay}\sim$1 Myr regardless of the stellar convective envelope's mass. Conversely, as the tidal efficiency gets lower, the effect of the star's convective envelope starts being more significant, and close to the lower theoretical limit $\koQs=2.516\times10^{-8}$ (see Fig. \ref{fig:WASPk2qdecay}) there is a more clear differentiation between the spiral-in time-scales produced by a low-mass (small $t\sub{decay})$ and high-mass convective envelope (large $t\sub{decay}$).

Fig. \ref{fig:WASPk2qdecay} represents the evolutionary history of WASP-19b for different characteristics of the system. In this Figure, leftmost and rightmost dots represent low- and high-mass convective envelopes, respectively (i.e. $\eps\gyrs$). Due to the spectral type of WASP-19, we assumed values of $\eps\gyrs$ between 0.001 and 0.1, as discussed by \citet{Dobs2004}. This range corresponds to stellar fluid convective envelopes which are significantly less massive than those assumed for NGTS-10, so the orbital angular momentum of WASP-19b is transferred to the rotation of WASP-19 at a much faster rate. For instance, to have $\eps\gyrs\simeq\gyrs$ (i.e. $\eps\simeq1$) means that the full convective envelope of the star is participating in the angular momentum exchange between the stellar rotation and planetary orbital motion, as assumed in the tidal evolution studied by \citet{Brown2011}. While such an assumption is qualitatively correct since either $\eps$ or $\gyrs$ are unknown quantities for WASP-19 (and generally most stars), it may give rise to some quantitative differences which are worthy to be addressed.

A three-dimensional parameter space  composed of $\alphas$ (17), $\betaso$ (17), and $\eps\gyrs$ (34) was used to analyse 9826 numerical simulations for the tidal evolution of WASP-19b, and find a relationship between the planet's orbital decay time-scale and the stellar TDR, oftentimes defined as the tidal quality factor (see e.g. \citealt{Ogilvie2004,Ogilvie2007,Jackson2008a}). This relation, which resembles an exponential decay, was found using an MCMC fit with a total of 300 walkers and 5000 steps, and corresponds to the green line in Fig. \ref{fig:WASPk2qdecay} where $a=1.13\times10
^{-6}$, $b=4.47$, and $c=5.53\times10^{-8}$. While this analysis is theoretical, it might help constrain WASP-19's tidal dissipation given some information about the spiral-in of WASP-19b. Put another way, if by other means (e.g. observationally) we were able to constrain the tidal decay time-scale of WASP-19b, we could obtain an approximate value of the tidal energy being dissipated by WASP-19, and from that value, which in the framework of this work is just an initial indicator, we could start to analyse the forward evolution of the system.

For the sake of illustrating and analysing the tidal evolution of WASP-19b, we adopted $t\sub{decay}$ halfway between the maximum and minimum values on the x-axis of Fig. \ref{fig:WASPk2qdecay} (i.e. $\sim$40.7 Myr), and used it in the aforementioned fitting to get the initial $\koQs$ of WASP-19 which corresponds to $\alphas=0.7$ and $\betaso=0.9$ in the bottom panel of Fig. \ref{fig:contours}. Then, once we set all of these parameters along with those mentioned at the beginning of this section, we let the system evolve for $\eps\gyrs$ ranging from 0.001 to 0.1. The semi-major axis evolution of WASP-19b is depicted in the bottom panel of Fig. \ref{fig:resultWASP}, where the foremost difference with the NGTS-10 system is the range of tidal decay time-scales which for WASP-19b are smaller ($\sim1.8-10$ Myr) when compared to NGTS-10b ($\sim10-60$ Myr). The main reason for this to happen is the eccentricity of WASP-19, which despite still being small, is considerably larger than that of NGTS-10b. Such a slightly more eccentric orbit would make WASP-19b interact more closely with its host star at some orbital points (near the periastron), producing a fast exchange of orbital and rotational angular momentum.

Additionally, we can see the evolution of WASP-19's $\koQs$ in the top-left panel of Fig. \ref{fig:resultWASP} happening at the same time frame than the semi-major axis and the planet's $\koQp$, the latter shown in the top-right panel of the same figure. As we can see, if WASP-19 is assumed to have a low-mass convective envelope, there exists a significant variation of $\koQs$ during the planet's spiral-in of more than an order of magnitude, going from a low- to a high-efficiency state of its tidal dissipation, and accelerating thereby the orbital decay of WASP-19b. On the contrary, when WASP-19 has a high-mass convective envelope, the evolution of $\koQs$ is negligible and the orbital decay takes longer. 

As shown in Fig. \ref{fig:finalperiodWASP}, the above has important effects on the final rotational periods of WASP-19 regarding different values of $\eps\gyrs$. We found that the star was significantly spun up for small $\eps\gyrs$ and final rotation periods were close to the stellar saturated regime (i.e. green region in Fig. \ref{fig:finalperiodWASP}), differing significantly from the initial state of the system and reaching values comparable to the planet's orbital mean motion (i.e. $\Os\lesssim\npp$). On the other hand, from Fig. \ref{fig:finalperiodWASP} it can be noticed that the larger $\eps\gyrs$ the larger the final stellar rotation period, meaning that a high-mass convective envelope made WASP-19 oppose the acceleration produced by the transfer of angular momentum from the planet's orbit. Strikingly, as we approach the right extreme value of $\eps\gyrs$, the evolution of the stellar rotation period seems to reach an asymptotic behaviour at approximately 9.8 d, which is fairly close to the initial rotation period of WASP-19; thus, for $\eps\gyrs>0.1$ the final rotational state of WASP-19 might not present any significant change, even though the planet would have transferred most of its orbital angular momentum. Still, for a G-type star like WASP-19, such values of $\eps\gyrs$ are unlikely as pointed out by \citet{Dobs2004}.

\section{Discussion}
\label{sec:dicussion}

For the sake of studying the evolution of tides in the orbital decay of ultra-short period (USP) planets, we limited our approach to the model detailed in Sections \ref{sec:evolution} and \ref{sec:tidal}. It is clear, however, that tidal dissipation efficiency is crucial when it comes to studying the orbital evolution of USP planets, and to neglect the evolution of stellar tides due to the possible spin-up produced by decaying {\it giant planets} (cf. \citealt{Hamer2020}) may lead to miscalculations of the time-scales required for planets to migrate inwards to their Roche limit, and eventually onto their host star. Nevertheless, a more accurate estimation of the internal energy dissipation in stars must include coupled effects with the stellar wind (magnetic braking), chemical differentiation of the fluid layers, differential rotation, and non-linear processes such as the turbulence in the fluid envelope. Indeed, the presence of double-diffusive convective layers and/or differential rotation could affect strongly the excitation, the propagation, and the dissipation of tidal waves (\citealt{Baruteau2013,Favier2014,Guenel2016}; \citealt*{Andre2017,Lin2018,Andre2019}).

In addition to the above, due to the significant stellar spin-up produced by the planet's orbital decay, magnetic effects on dynamic tides become important when explaining the tidal dissipation history of the system \citep{Wei2016,Wei2018,Astoul2019}. During orbital decay, tidal heating produced by the dissipation of tides inside the planet might also modify its interior. Despite this would not significantly alter either the global angular momentum exchange or the planet's final fate, it can be used to improve planetary internal structure and gravity-field models (see e.g. \citealt*{Militzer2019,Debras2019}). Interestingly, this phenomenon could also have effects even for natural satellite systems orbiting the planet at its final position, where effects such as \textit{Resonance Locking} may appear \citep{Fuller2016,Lainey2020}.

It is worth noting that equations (\ref{eq:k2QFormulas}) and (\ref{eq:k2QFormulap}) are average values of different tidal frequencies upon which the dissipation properties of a spherical shell may depend \citep{Ogilvie2007}, so we could be ruling out different effects arising from such dependence and thereby under/overestimating the energy dissipation in the stars. Moreover, all the expressions in Section \ref{sec:tidal} neglect the friction in the fluid-solid or fluid-fluid layers, albeit this frictional force might exist and be caused by differential rotation producing a frequency-lag in the different layers. Also, the dependence on the internal viscosity has been removed for the sake of simplicity. A more refined model for the dissipation of energy in solid cores should take such an effect into account.

Other works that have included stellar evolution into the orbital evolution of planets \citep{Penev2014}, have mainly studied planets which are stable against orbital decay and kept $\koQp$ and $\koQs$ constant in time. Authors like \citet{Bolmont2016} and \citet{Benbakoura2019} have analysed the coupling between tidal dissipation and effects arising from the properties of the convective stellar envelope, taking into account stellar structural changes and wind braking. In this work, we explored some of these effects along with those produced in systems where giant planets are undergoing orbital decay and  transferring angular momentum to the stellar spin. From the results we came to the conclusion that a dynamical approach for $k_2/Q$ for both the star and the planet is needed when studying the orbital evolution of USP planets; then, no constant phase/time lag model of the tidal bulge was used. In this regard, the system of evolutionary equations presented in Section \ref{sec:tidal} comes from the application of the work of \citet{Hut1981}, with no truncation in eccentricity to prevent any incorrect computation of the tidal evolution of the studied systems; although such problems are more prone to appear when eccentricities are larger than 0.2, as explained by \citet{Leconte2010}.

Given the proximity of USP planets to their host star, there are different effects related to physical changes that were also taken into account, including but not limited to planet's mass loss, size contraction, and changes in the interior structure through the planetary Tidal Dissipation Reservoir (TDR), $\koQp$. However, the orbital decay time-scales of USP planets such as NGTS-10b and WASP-19b are only of the order of Myr, having reached asymptotic characteristics whose variations take much longer than orbital decay. We showed here that the evolving tidal dissipation within the star is the dominant mechanism directing the transfer of angular momentum in the systems NGTS-10 and WASP-19, and that the study of exoplanet tidal evolution should include the existing connection between the star's rotation, magnetic braking, stellar mass loss rate, and the periodical tidal dissipated energy via $\koQs$.

%%%%%%%%%%%%%%%%%%%%%%%%%%%%%%%%%%%%%%%%%%%%%%%%%%%%%%%%%%%%%%%%%%%%%%
% FIGURE
%%%%%%%%%%%%%%%%%%%%%%%%%%%%%%%%%%%%%%%%%%%%%%%%%%%%%%%%%%%%%%%%%%%%%%
\begin{figure}
        \hspace{-9pt}
         \includegraphics[scale=0.61]{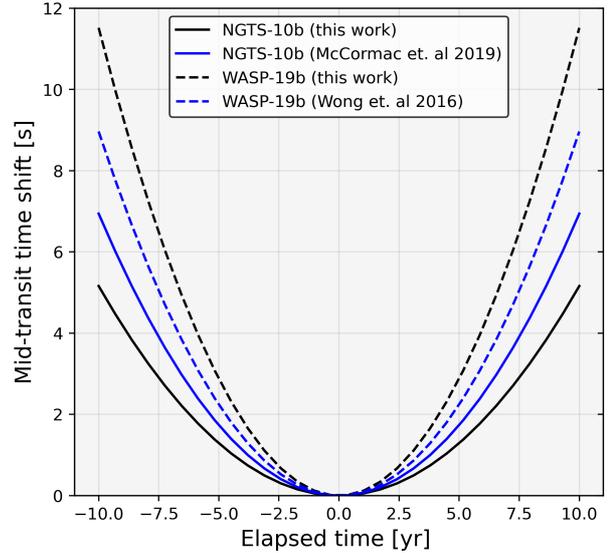}
    \caption{Predicted shift in mid-transit times for NGTS-10 (solid line) and WASP-19 (dashed line). The black lines correspond to this work, while blue lines are those computed by \citet{McCormac2019} for NGTS-10, and by \citet{Wong2016} for WASP-19.}
\label{fig:humandeltas}
\end{figure}
%%%%%%%%%%%%%%%%%%%%%%%%%%%%%%%%%%%%%%%%%%%%%%%%%%%%%%%%%%%%%%%%%%%%%%%

Stellar age is directly related to the time-scales of orbital decay and the spin-up of the star. If the stars are older than assumed, orbital decay is still dominant and the time-scales found here keep approximately the same, but if we chose the lower limit for the age of WASP-19\footnote{We exclude NGTS-10 from this analysis as it is markedly old, so using either the upper or lower bound of its age is negligible for the orbital decay of the planet and the spin-up of the star.}, orbital decay is considerably delayed and the star spins up at a slower rate. This occurs because stellar wind transfers large amounts of angular momentum from the stellar spin to the planetary orbit. In extreme cases (not shown in this work), when stars are too young the orbital decay of USP planets is not enough to spin up the stellar rotation and magnetic braking becomes the dominant mechanism driving tidal evolution. As a result, the star spins down instead of spinning up. Similarly, this will also happen once USP planets have reached their Roche limit and stars are left with high rotational rates that reactivate their wind. However, as can be seen in Figs \ref{fig:omegaspin}, \ref{fig:finalperiodNGTS}, and \ref{fig:finalperiodWASP} this effect is more significant if either NGTS-10 or WASP-19 have a low-mass convective envelope.

Since stellar ages in this work were adopted from previously computed values in the literature, the results presented here may have an impact to constrain the upper, but mainly the lower age bound of the considered systems. As expected, stellar age, spin, orbital decay time-scales, and tidal energy are all entwined in such a way that by solving for one of them we can shed light on the remaining. Having said that, this work may be used to constrain $\koQs$ and derive stellar rotation periods, which in turn would allow us to estimate stellar ages. For WASP-19, our upper limit for $\koQs$ (see Section \ref{sec:wasp19b}) gives a constraint of $\tage >1$ Gyr, similar to that found by \cite{Hansen2010} and discussed by \cite{Brown2011}; with corresponding orbital decay time-scales $t\sub{decay}\lesssim3$ Myr. However, the lower limit proposed here for $\koQs$ (see Section \ref{sec:wasp19b}) seems to suggest an older WASP-19 of $\tage > 7$ Gyr in agreement with recent stellar model fits (e.g. \citealt{Demarque2004,Marigo2008}), and recent analysis of the spectral lithium abundance \citep{Brown2011}. In such scenarios, the orbital decay time-scales of WASP-19b would be larger than 10 Myr.

Despite the purpose of this work is not to compute accurate stellar ages, a well defined step-by-step guide could be followed hereafter to further constrain $\tage$. That is, speaking in human time-scales, if we succeeded in the long-term to measure the mid-transit time shifts of either WASP-19b or NGTS-10b, we may recalculate the planetary orbital decay time-scales and thus estimate $\koQs$, stellar rotation periods, and hence the stellar age. This process is summarised in Fig. \ref{fig:WASPk2qdecay}. For NGTS-10 an upper limit of $\koQs\approx10^{-7}$ was found for a corresponding orbital decay time-scale of 5 - 60 Myr. This estimate agrees with a stellar age larger than 7 Gyr, which agrees with the stellar model fit performed by \citet{McCormac2019}. More observations are needed to further constrain the age of NGTS-10.

Star-(USP)planet tidal interactions spin up the host star while the planet is moving towards the Roche limit, where  planets may continue their spiral-in until eventually being gobbled down by the star's surface. Planets could also be disrupted by tidal forces before falling onto the star, leaving a mix of debris that might lead to other planetary phenomena. An interesting scenario might be the possibility for the star to reach a rotational rate capable of producing a reversal on the orbital evolution of the planet making it migrate outwards. However, we have limited our approach to purely dynamical simulations and no hydrodynamical effects were taken into account, so the forward evolution once NGTS-10b and WASP-19b reached their Roche limit was out of the scope of this work. 

Since the discovery of WASP-19b by \citet{Hebb2010}, this system has been under both theoretical and observational thorough examination. From theory we expect WASP-19b to undergo orbital decay produced by the tidal interactions with its host star, similar to what would happen with most close-in giant planets; thus, authors such as \citet{Valsecchi2014} and \citet{Essick2016} have predicted that owing to WASP-19's decreasing orbital period, observational human-scale changes of its mid-transit times could be detected, and such predictions may be theoretically assessed using formalisms like that proposed by \citet{Cameron2018}. Still, dedicated campaigns with the highest photometric quality and fast-cadence observations are needed to measure (in human time-scales) an astronomical process that might take millions of years.

Recently, after analysing a 10-year baseline of WASP-19b's observational data, \citet{Petrucci2020} found no evident shift of its mid-transit times, casting doubt on its orbital decay. However, tidal dissipation efficiency as parametrized through $\koQs$ is key for understanding the orbital evolution of WASP-19b and whether specific conditions are required to trigger the spiral-in of the planet. As mentioned throughout this work, $\koQs$ is crucial to study the relation between the star's convective mass and the tides raised by the planet on the star: to shed some light on how such entangled parameters affect the planet's orbital decay that may or may not produce measurable changes in transit timing, we need to grasp the mechanisms driving the tidal evolution of compact systems. Thus, this work is intended as an attempt to help conciliate different effects concurring in the same phenomenon; however, orbital decay is still a puzzle which also needs to be supported via observation.

Finally, a careful analysis of the transit light curves of NGTS-10b and WASP-19b could lead to unveiling small variations of transit shape, depth, and duration produced by the large deformations of the circular shape of close-in planets as a result of strong stellar tides (see e.g. \citealt{Correia2014, Akinsanmi2019}). The observable property that could arguably be the most affected is the planetary mid-transit time since orbital decay means smaller orbital periods as time goes on, making the transits of NGTS-10b and/or WASP-19b come early if we considered a large enough baseline for an observing campaign. By following \citealt{Cameron2018}, for an elapsed time of 10 yr, we predict a mid-transit time advance of 5 and 11 s for NGTS-10b and WASP-19b, respectively (see Fig. \ref{fig:humandeltas}).

\vspace{-0.2cm}
\section{Conclusive remarks}
\label{sec:conclusion}

We presented a general model that can be used for studying the tidal evolution of USP giant planets. Such a model couples many different effects taking place in star-planet systems that modify the duration of what is formally known as orbital decay. Using this model we studied the orbits of NGTS-10b and WASP-19b, the two shortest-period planets known to date, and found that their spiral-in process strongly depends on the interior characteristics of their host star, namely the tidal dissipation efficiency and the convective fluid envelope's mass.

After the analysis of thousands of numerical simulations, we proposed a theoretical relation between the orbital decay time-scale of USP planets and the tidal dissipation reservoir of their host star. By following the same method such a relation can be replicated for other extrasolar systems with USP planets (see Fig. \ref{fig:distrib}) such as KELT-16, HATS-18, HIP 65 A, NGTS-6, WASP-43, and WASP-103; and by constraining orbital decay time-scales we could have an approximate value of the dissipation of energy inside the star. Although there are still many different facets of compact systems and their internal processes that entail additional research work, any attempt to delve into the mechanisms behind star-planet tidal interactions is a potential step forward to finding observational evidence of tidal evolution theories.

\section*{Acknowledgements}

The authors thank the anonymous referee for the helpful feedback and positive remarks, which allowed us to improve the quality of this research. JAA-M acknowledges funding support from Macquarie University through the International Macquarie University Research Excellence Scholarship (`iMQRES'). MS acknowledges support by ANID, -- Millennium Science Initiative Program -- NCN19\_171.
and Colciencias. This research has made use of the NASA's Astrophysics Data System (ADS) and NASA Exoplanet Archive, which is operated by the California Institute of Technology, under contract with the National Aeronautics and Space Administration under the Exoplanet Exploration Program.

\section*{Data Availability}

The data underlying this article will be shared on reasonable request to the corresponding author.

%%%%%%%%%%%%%%%%%%%%%%%%%%%%%%%%%%%%%%%%%%%%%%%%%%

%%%%%%%%%%%%%%%%%%%% REFERENCES %%%%%%%%%%%%%%%%%%

% The best way to enter references is to use BibTeX:

%\bibliographystyle{mnras}
%\bibliography{references} % if your bibtex file is called example.bib
\input{thespiralin.bbl}

%%%%%%%%%%%%%%%%%%%%%%%%%%%%%%%%%%%%%%%%%%%%%%%%%%
%%%%%%%%%%%%%%%%%%%%%%%%%%%%%%%%%%%%%%%%%%%%%%%%%%

% Don't change these lines
\bsp	% typesetting comment
\label{lastpage}
\end{document}